\documentclass[9pt,journal,compsoc]{IEEEtran}
  \usepackage{cite}
\usepackage{caption}
\usepackage{cleveref}
\usepackage{algpseudocode}
\usepackage[]{algorithm2e}
\usepackage[pdftex]{graphicx}
\ifCLASSINFOpdf
\else
\fi
\usepackage{url}

\begin{document}
%
\title{Memory-Efficient Quantum Circuit Simulation by Using Lossy Data Compression}
%
%
%
%

\author{Xin-Chuan~Wu,~\IEEEmembership{}
        Sheng~Di,~\IEEEmembership{}
        Franck~Cappello,~\IEEEmembership{}
        Hal~Finkel,~\IEEEmembership{}
        Yuri~Alexeev,~\IEEEmembership{}
        and~Frederic~T.~Chong ~\IEEEmembership{}
\IEEEcompsocitemizethanks{\IEEEcompsocthanksitem Xin-Chuan Wu is with Department of Computer Science, University of Chicago, Chicago, IL 60637. E-mail: xinchuan@uchicago.edu
\IEEEcompsocthanksitem Sheng~Di is with Argonne National Laboratory, Lemont, IL 60439.
\IEEEcompsocthanksitem  Franck~Cappello is with Argonne National Laboratory and University of Illinois Urbana-Champaign, Urbana, IL 61801.
\IEEEcompsocthanksitem Hal~Finkel and Yuri~Alexeev are with Leadership Computing Facility, Argonne National Laboratory, Lemont, IL 60439.
\IEEEcompsocthanksitem Frederic~T.~Chong is with Department of Computer Science, University of Chicago, Chicago, IL 60637.}
\thanks{PMES Workshop, Dallas, 11 Nov 2018. \texttt{http://j.mp/pmes18}}}

%
%

\markboth{3rd International Workshop on Post-Moore's Era Supercomputing (PMES)}%
{PMES Workshop, Dallas, 11 Nov 2018}
%



\IEEEtitleabstractindextext{%
\begin{abstract}
In order to evaluate, validate, and refine the design of new quantum algorithms or quantum computers, researchers and developers need methods to assess their correctness and fidelity. This requires the capabilities of quantum circuit simulations. However, the number of quantum state amplitudes increases exponentially with the number of qubits, leading to the exponential growth of the memory requirement for the simulations.
In this work, we present our memory-efficient quantum circuit simulation by using lossy data compression.
Our empirical data shows that we reduce the memory requirement to 16.5\% and 2.24E-06 of the original requirement for QFT and Grover's search, respectively. This finding further suggests that we can simulate deep quantum circuits up to 63 qubits with 0.8 petabytes memory.
\end{abstract}
}


\maketitle

\IEEEdisplaynontitleabstractindextext

%
\IEEEpeerreviewmaketitle

 \section{Introduction}\label{sec:intro}

Using classical computing systems to simulate quantum computers is important for better understanding the operations and behaviors of quantum computing systems. Such simulations allow researchers to evaluate the complexity of new quantum algorithms and validate the design of quantum circuits.  Previous studies have provided different techniques, such as full amplitude-vector update \cite{de2007massively, smelyanskiy2016qhipster, haner20170}, Feynman paths \cite{bernstein1997quantum}, and tensor network contractions \cite{markov2008simulating, pednault2017breaking, boixo2017simulation, chen201864, chen2018classical} to perform the simulation of quantum circuits. Full amplitude-vector update simulations provide all amplitudes of the quantum states in detail, and hence it is the best tool for quantum algorithms debugging, development, and validation. In full amplitude-vector update simulations, we generally use complex double precision amplitudes to represent the state of the quantum systems. Given $n$ quantum bits (qubits), we need $2^n$ amplitudes to describe the quantum system. The state vector requires $2^{n+4}$ bytes. Since the number of quantum state amplitudes grows exponentially with the number of qubits in the system, the size of the quantum circuits simulation is limited by the memory capacity of the classical computing system. For example, to store the full quantum state of a 45-qubit system, the memory requirement is 0.5 petabytes. Circuits with more than 49-qubit system would require too much memory to simulate.

Since the quantum circuits simulation size is restricted by the memory capacity of the classical computing systems, our goal is to trade the computation time for the memory space. We compress the state vector so that we can have a larger memory space for more state vector, and hence we can simulate a larger quantum system. The performance overhead is expected, but we are able to simulate a larger quantum system with the limited memory capacity. In this work, we incorporate HPC lossy data compression techniques to the quantum circuits simulator, Intel-QS \cite{smelyanskiy2016qhipster}. Intel-QS is a distributed high performance quantum circuits simulator. Since Intel-QS uses full state amplitude-vector update technique to run the simulation, Intel-QS is capable of high depth quantum circuits simulation.

By using data compression to reduce the memory requirement of storing the full quantum state, we are able to simulate larger quantum systems within the same memory capacity. The scalability of this approach is determined by the quantum state vector compression ratio. Compared to lossless compression, lossy compression algorithms achieve more significant compression ratios in general. Thus, we employ the SZ lossy compressor \cite{xin2018, tao2017significantly, di2016fast} to our memory-efficient quantum circuits simulation framework.

\section{Methodology}

\begin{algorithm}[htbp]
 \For{each gate G in the program}{
  \For{each slice S in the state vector}{
   Decompress(S)\\
   Normalize(S)\\
   GateComputation(G, S)\\
   Compress(S)\\
  }
 }
 \caption{Quantum state vector slice update.}
 \label{algo:operations}
\end{algorithm}

In our simulation, we store all the quantum state amplitudes as a complex vector. The full quantum state vector is divided into several slices. Each slice is compressed and stored on memory. Only the slice under processing will be decompressed. The pseudo-code of the implementation is shown in Algorithm~\ref{algo:operations}. The simulation of a quantum program is processed gate by gate. To apply a quantum gate, each slice must be decompressed. Since lossy compression introduces compression errors into the state vector, the state vector is no longer located exactly on the block sphere \cite{nielsen2002quantum}. To reduce the error propagation, the state vector is normalized before the gate computation. After the normalization, we apply the gate computation to the vector slice, and then compress the vector slice. This is a complete operation cycle for a slice. After a slice is finished, we process the next slice.

SZ lossy compressor allows the user to set the error bound. To maintain the fidelity of the state, we set the error bound for each amplitude number to 1\%. However, if one wants to have a higher compression ratio, it is flexible to trade fidelity for compression ratio. In addition, it is intuitive that if a dataset contains a lot of zeros or the same values, we can easily get a high compression ratio, and if a dataset contains various values, we might only get a limited compression ratio.

To improve the performance of the simulation, we utilize Message-Passing Interface (MPI) \cite{gropp1999using, gropp1996high} to process multiple slices in parallel.

\section{Results}\label{sec:results}

To provide the proof of concept, we developed and performed our memory-efficient quantum circuits simulation on Argonne Theta supercomputer.  We evaluated our approach with two quantum algorithms, (1) Quantum Fourier Transform (QFT), and (2) Grover's search algorithm. QFT is one of the most popular quantum applications, and Grover's search algorithm is one of the most famous quantum algorithms. We assess the simulation quality by the state fidelity, the compression ratio, and the simulation time. Fidelity is a measure of the similarity of two quantum states \cite{nielsen2002quantum}. If the fidelity value is 1, then the two quantum states are identical. The compression ratio determines how many extra qubits we can simulate. The simulation time reflects how many gates we can simulate in a time period.

In QFT experiment, we run the simulation with different size of the QFT benchmarks, shown in Table~\ref{tab:qft}. Since the compression ratio is correlated to the original data values, and QFT generates different amount of amplitudes, this test case stands for the worst case scenario for the compression ratio.

\begin{table}[h]
\centering
\begin{tabular}{c| c c}
 & \# of qubits & \# of gates \\\hline\hline
QFT 20 & 20 & 1012 \\\hline
QFT 26 & 26 & 1732 \\\hline
QFT 30 & 30 & 2270
\end{tabular}
\caption{QFT benchmarks}
\label{tab:qft}
\end{table}

\begin{figure}[!t]
\centering
\captionsetup{justification=centering}
\includegraphics[width=0.4\textwidth,keepaspectratio]{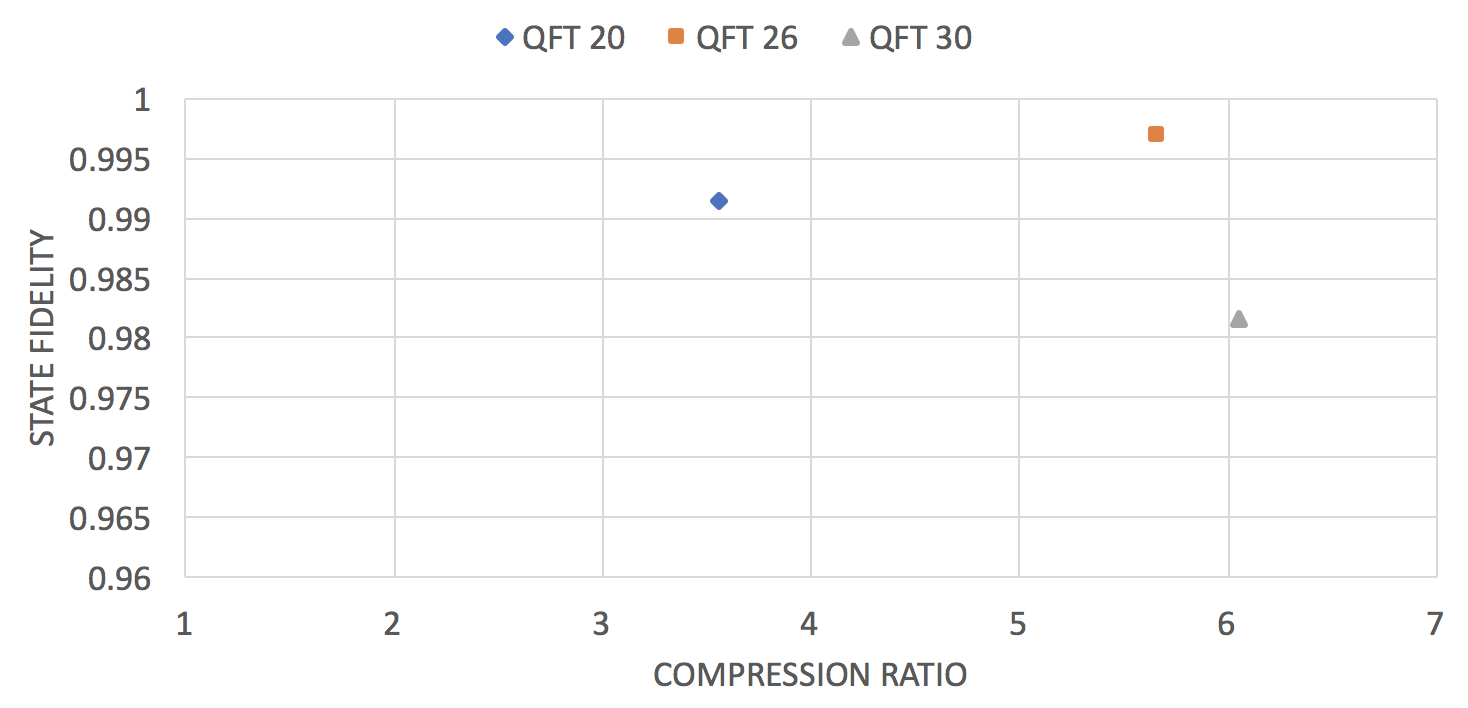}
\caption{Compression ratio and fidelity of QFT benchmarks.}
\label{fig:qft}
\end{figure}

Figure~\ref{fig:qft} shows the minimum compression ratio and fidelity of three QFT benchmarks. We care about the minimum compression ratio because this limits the number of extra qubits we can simulate. Fidelity is a measure of the similarity of two quantum states \cite{nielsen2002quantum}. If the fidelity value is 1, then the two quantum states are identical. We calculate the state fidelity of the final amplitudes vectors between our approach and the original simulator. The result shows that our approach can achieve 98\% fidelity and the compression ratio for 30 qubits system is greater than 6x. This compression ratio suggests that we can simulate 2 more qubits with the current memory capacity.

\begin{figure}[!t]
\centering
\captionsetup{justification=centering}
\includegraphics[width=0.4\textwidth,keepaspectratio]{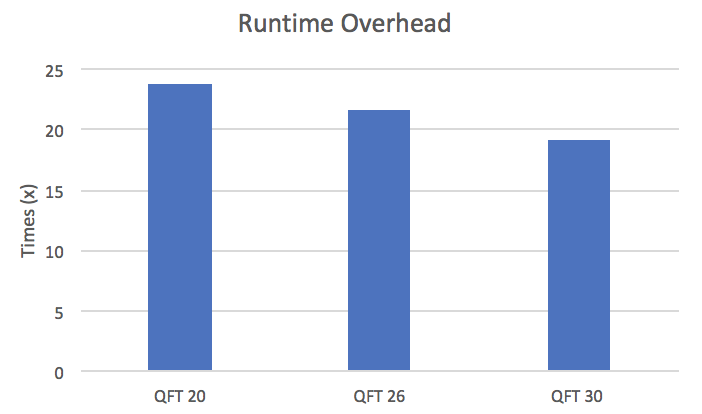}
\caption{Performance overhead of QFT benchmarks.}
\label{fig:overhead}
\end{figure}

We expect to have performance overhead since we introduce compression, decompression, and normalization processes into our simulation. Figure~\ref{fig:overhead} shows the performance overhead. Our memory-efficient simulation run-time overheads are 24x, 21x, and 19x  for QFT20, QFT26, and QFT30, respectively.

In the Grover's search algorithm, most of the state amplitudes are the same value during the simulation. Such state vectors allow the SZ compressor to perform high compression ratios. In our Grover's search benchmark (30 qubits), the minimum compression ratio is 445144. We achieve 99.75\% fidelity. The simulation time is 19x in this experiment.

The supercomputer, Theta, is able to simulate a 45-qubit quantum system with 0.8 petabytes memory. Our experimental results suggest that we can compress the state vector to get memory space to add 18 ($\lfloor{log_2445144}\rfloor$) qubits for Grover's search algorithms simulation. Thus, our ongoing work is to implement the simulation of 63-qubit Grover's search algorithms on Theta. 

\section{Discussion}\label{sec:disscussion}

To evaluate quantum algorithms, the simulation size must be greater than the number of qubits used in the algorithms. In the previous work, it is impossible for full amplitude vector update simulators to validate the quantum algorithms using more than 49 qubits. In this paper, we propose a memory-efficient quantum circuits simulation technique to simulate more qubits than previously reported by using lossy data compression. We trade computation power for memory space, and we further trade fidelity for high compression ratio, so that we can safely expect to achieve the simulation of quantum circuits beyond 50 qubits with 0.8 petabytes.

According to our experimental results, we can get 6x compression ratio with 19x performance overhead in QFT30, which means we can increase 2 qubits to our simulation with the same memory capacity. In fact, the trend in Figure~\ref{fig:qft} and Figure~\ref{fig:overhead} imply that we should get a higher compression ratio with lower performance overhead if we increase our simulation size to 45 qubits. This is because the portion of the compression overhead is relatively small when the dataset is a large data. In the Grover's search case, which is in favor of our approach, we can get 445144 compression ratio, which is equivalent to 18 qubits increment in our simulation. The reason we get high compression ratio is that most of the amplitudes in this algorithm have similar values. Since we can find this characteristics in several quantum algorithms, we expect the similar compression ratio can be achieved for simulating various quantum algorithms.

In the future work, we plan to run 63-qubit Grover's search algorithm simulation on Theta, analyze the effect of compression errors and relationships to real physical noise, integrate our technique with other approximate simulation techniques, and evaluate different compression algorithms for quantum state amplitudes.

\clearpage
\appendices

\section*{Acknowledgments}
This research used resources of the Argonne Leadership Computing Facility, which is a DOE Office of Science User Facility supported under Contract DE-AC02-06CH11357.This research was supported by the Exascale Computing Project (ECP), Project Number: 17-SC-20-SC, a collaborative effort of two DOE organizations – the Office of Science and the National Nuclear Security Administration, responsible for the planning and preparation of a capable exascale ecosystem, including software, applications, hardware, advanced system engineering and early testbed platforms, to support the nation’s exascale computing imperative. The material was supported by the U.S. Department of Energy, Office of Science, and supported by the National Science Foundation under Grant No. 1619253. This work is funded in part by EPiQC, an NSF Expedition in Computing, under grant CCF-1730449. This work is also funded in part by NSF PHY-1818914 and a research gift from Intel.



\bibliographystyle{./IEEEtranS}
\bibliography{./refs}

\end{document}